\documentclass[%
 reprint,
 amsmath,amssymb,
 aps,
 prb,
]{revtex4-1}

\usepackage{graphicx}
\usepackage{dcolumn}
\usepackage{bm}

\newcommand{\calH}{\mathcal{H}}
\newcommand{\calO}{\mathcal{O}}

\newcommand{\calJ}{\mathcal{J}}
\newcommand{\eff}{\mathrm{eff}}
\usepackage{braket}
\begin{document}

\preprint{APS/123-QED}

\title{Laser-induced topological superconductivity in cuprate thin films  }

\author{Kazuaki Takasan}
\email{takasan@scphys.kyoto-u.ac.jp}
\author{Akito Daido}
\author{Norio Kawakami}
\author{Youichi Yanase}
\affiliation{%
 Department of Physics, Kyoto University, Kyoto 606-8502, Japan
}%

\date{\today}

\begin{abstract}
We propose a possible way to realize topological superconductivity with application of laser light to superconducting cuprate thin films. 
Applying Floquet theory to a model of $d$-wave superconductors with Rashba spin-orbit coupling, we derive an effective model and discuss its topological nature. Interplay of the Rashba spin-orbit coupling and the laser light effect induces the synthetic magnetic fields, thus making the system gapped. Then the system acquires the topologically non-trivial nature which is characterized by Chern numbers. The effective magnetic fields do not create the vortices in superconductors, and thus the proposed scheme provides a promising way to dynamically realize a topological superconductor in cuprates. We also discuss an experimental way to detect the signature.
\end{abstract}

\maketitle


\section{Introduction}
Topological superconductors (TSCs) have attracted a great deal of interest 
from the viewpoint of realization of Majorana fermions in solid states and a possible application to quantum computation \cite{Qi2011, Sato2016}.
However, it is the current situation that experimental realization of TSCs is still limited.
There are two main ways to search for topological superconductors.
First one is to engineer a TSC in artificial systems by proximity effect (artificial TSCs).  
In recent studies, there are substantial developments in some artificial systems, such as a ferromagnetic atomic chain on a superconductor \cite{Nadj-Perge2014} or a nanowire on a superconductor \cite{Mourik2012}. 
In these artificial systems, the realization of TSCs is confirmed to a certain extent. 
Second way is to find a material which is intrinsically a TSC (intrinsic TSCs).
Some candidate materials of intrinsic TSCs have been proposed, for instance, spin-triplet superconductor $\mathrm{Sr_2RuO_4}$ \cite{Kashiwaya2011} 
and doped topological insulators $\mathrm{Cu_xBi_2Se_3}$ \cite{Sasaki2011, Matano2016, Yonezawa2016}. 
However, nodal excitation in nearly gapless $\mathrm{Sr_2RuO_4}$
\cite{Nishizaki2000, Ishida2000} is harmful for experimental detection
of topological response, and topological nontriviality  in
$\mathrm{Cu_xBi_2Se_3}$
is still under debate \cite{Levy2013}.
Therefore, further search of intrinsic TSCs is one of the important issues in the research field.

On the other hand,  in recent years, tremendous developments have been achieved in the realization of topological phases by laser light applications\cite{Oka2009, Lindner2011, Grushin2014, Titum2015}.
A typical example of the laser-induced topological state is a quantum Hall state in graphene \cite{Oka2009}. 
This state is induced not by static magnetic field but by circularly polarized dynamical laser light.
In this case, laser light effectively induces the next-nearest hopping with complex phase, which makes the system gapped, 
and thus the topologically non-trivial states similar to Haldane model \cite{Haldane1988} are realized.
Though this phenomenon has not been confirmed experimentally in graphene, 
a similar phenomenon is observed on the surface of the laser-irradiated topological insulators by time-resolved ARPES (Tr-ARPES) experiments \cite{Wang2013}.
The ARPES image obtained in the experiments shows that the surface Dirac cone becomes gapped when the laser light is applied to the system. 
Thus, we can regard the laser light as a new tool for controlling the topology of the states of matter.

Motivated by these situations, in this paper, we propose a possible way to realize TSCs with application of the laser light to well-known materials. 
We discover that the TSCs can be realized in $d$-wave superconductors, such as cuprate, fabricated on a substrate irradiated by circularly polarized laser light.
We apply Floquet theory to a model of $d$-wave superconductor and derive an effective model under the irradiation of the laser light.
Based on this effective model, we reveal that the system acquires the topologically nontrivial nature, which is characterized by Chern numbers, and show that the laser-induced magnetic field in the effective model play a crucial role in realizing TSCs. 

This paper is organized as follows.
In Sec. II, we introduce our model and methods. 
Next we show the derivation of an effective model which describes laser-irradiated cuprate thin films in Sec. III.
In Sec. IV, we discuss topological properties of the effective model.
We show the topological phase diagram and clarify the nature of each phase.
In Sec.V, we discuss the experimental conditions to realize TSCs.
Finally, summary and outlook are presented in Sec. VI.

\section{Model and Methods}

\begin{figure}[tbp]
\includegraphics[width=7.5cm]{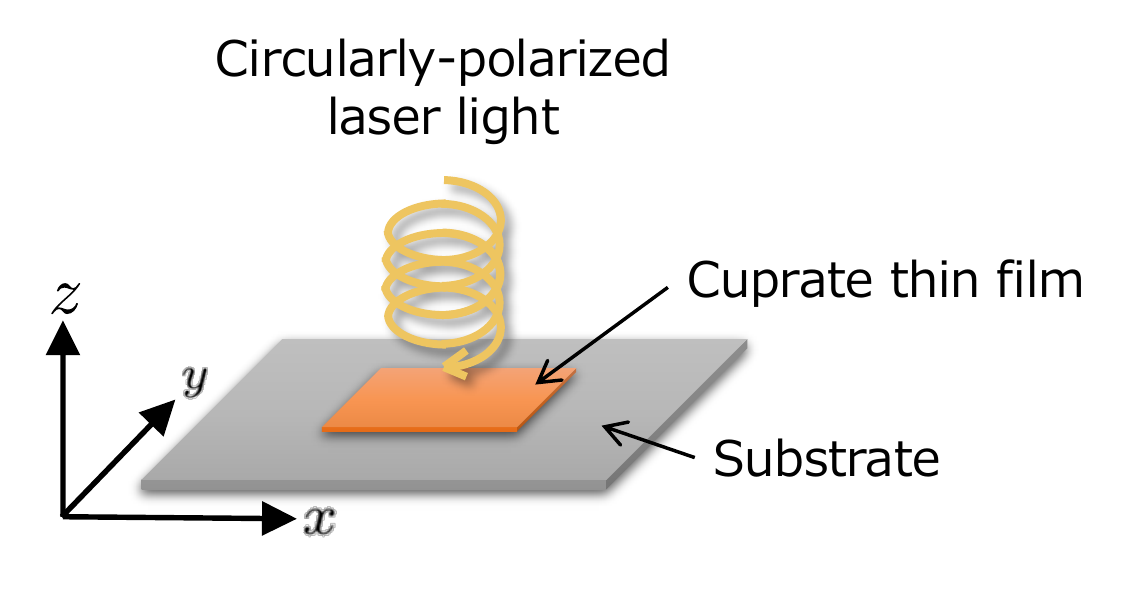}
\caption{Schematic picture of the setup. A cuprate thin film fabricated on a substrate is irradiated by circularly polarized laser light in the $z$-direction.}
\label{setup}
\end{figure}

A setup of the system is schematically shown in Fig. \ref{setup}. 
We consider a cuprate thin film fabricated on a substrate. 
Because of the asymmetric potential due to the substrate, Rashba spin-orbit coupling appears. 
In order to describe such a situation, we introduce a Rashba-Hubbard model as 
\begin{align}
\calH&=\sum_{\bm k \sigma} \xi (\bm k) c_{\bm k \sigma} ^\dagger c_{\bm k \sigma} 
\nonumber \\
&+\sum_{\bm k \sigma \sigma'} (\alpha \bm g (\bm k) \cdot \bm \sigma)_{\sigma \sigma'}
c_{\bm k \sigma} ^\dagger c_{\bm k \sigma'} 
+U \sum_i n_{i \uparrow} n_{i \downarrow}, 
\end{align}
where
\begin{align}
\xi ({\bm k})&=-2 t(\cos k_x+\cos k_y) + 4t^\prime \cos k_x \cos k_y - \mu, \\
\bm g (\bm k)&=(-\sin k_y, \sin k_x, 0),
\end{align}
with $ c_{\bm k \sigma} $ being the annihilation operator of electrons with momentum $\bm k$ and spin $\sigma$.  We choose the form of $\xi ({\bm k})$, which includes the next-nearest neighbor hopping $t^\prime$, in order to reproduce the Fermi surface of typical high-$T_c$ cuprates.

Next we consider the effect of laser light. 
We treat the laser light as time-dependent classical electromagnetic fields $\bm A(t)$ and introduce them as Peierls phases. 
This treatment is equivalent to substituting $\bm k $ with $ \bm k - \bm A(t)$. With this substitution, we obtain the time-dependent model, which describes
laser-illuminated cuprate thin films, as
\begin{align}
\calH(t)&=\sum_{\bm k \sigma} \xi (\bm k - \bm A (t)) c_{\bm k \sigma} ^\dagger c_{\bm k \sigma} 
\nonumber \\
&+\sum_{\bm k \sigma \sigma'} (\alpha \bm g (\bm k- \bm A (t)) \cdot \bm \sigma)_{\sigma \sigma'}
c_{\bm k \sigma} ^\dagger c_{\bm k \sigma'} 
+U \sum_i n_{i \uparrow} n_{i \downarrow} \label{t-dep}
\end{align}
with $\bm A(t) = (A_x \cos \omega t,A_y \sin \omega t ,0)$, which corresponds to circularly ($A_x=A_y$) or elliptically ($A_x \neq A_y$) polarized laser light. 
Generally speaking, it is difficult to solve time-dependent quantum many-body problems. 
However, as for time-periodic problems, we can solve them by Floquet theory, which is known as a useful tool for analyzing time-periodic systems.
The model Hamiltonian (\ref{t-dep}) is periodic in time and therefore we can apply Floquet theory to it. 

Floquet theory is based on Floquet theorem, which is, so to speak, ``Bloch's theorem for time direction'': 
If the Hamiltonian is time-periodic $\calH(t)=\calH(t+T)$, 
the eigenfunction can be written by a product of an exponential function $e^{-i\epsilon t}$ and a time periodic function $u(t)$. $\epsilon$ is called ``pesudo energy'' that is defined in the range of $-\pi/T < \epsilon < \pi/T=\omega/2$. 
Then, we can define the effective Hamiltonian, of which eigenvalues are pesudo energy, as
\begin{align}
\calH_{\eff}=\frac{i}{T}\log U(T),\label{Heff_def}
\end{align} 
where the time-evolution operator $U(t)=\mathcal{T}\exp \left(-i \int_0^{t} \calH(s)ds \right) $. 
By definition, the effective Hamiltonian has only the information of $t=0,T,2T,\cdots, nT,\cdots$ and gives ``stroboscopic description" of this system. 
In other words, the information of the time range of $nT<t<(n+1)T$ is neglected. 
Intuitively, if the time period is short enough, this Hamiltonian tells us the asymptotic behavior of periodically-driven systems.
Although it is not obvious whether the effective Hamiltonian fully describes the quantum states while the system is driven \cite{DAlessio2014}, 
it is recently shown for many-body quantum systems that nonequilibrium steady states can appear with finite life time when the frequency is sufficiently high\cite{Kuwahara2016}. It is also shown that the nonequilibrium steady states are described by an effective Hamiltonian in Floquet theory\cite{Kuwahara2016}. 
Thus we can use the effective Hamiltonian for understanding the nature of the non-equilibrium steady states of laser-irradiated materials including strongly correlated electron systems.

Though it is difficult to directly calculate the effective Hamiltonian from the definition (\ref{Heff_def}), there are some useful methods to derive the effective Hamiltonian.
We adopt a perturbative expansion in $1 / \omega$  \cite{Mikami2016}. Using this approach, we can write down the effective Hamiltonian as
\begin{align}
\calH_{\mathrm{eff}} = \calH_0 + \sum_n \frac{[\calH_{-n},\calH_n]}{n \omega}+\calO\left(\omega^{-2}\right), \label{Heff_expansion}
\end{align}
where $\calH_n =\frac{1}{T} \int^{T/2}_{-T/2}dt \calH(t) e^{-i n \omega t}$. The second term in the case of the laser-irradiated systems represents the second order perturbation process of the $n$-photon absorption $\calH_n$ and $n$-photon emission $\calH_{-n}$ in off-resonant light. If the laser intensity is sufficiently small, it is reduced to virtual processes related to one-photon absorption $\calH_1$ and emission $\calH_{-1}$.

\section{Derivation of the effective model}

In this section, we calculate the effective Hamiltonian (\ref{Heff_expansion}) up to the first order in  $1 / \omega$. The effective model describes the nonequilibrium steady states of the irradiated cuprate thin film. The model is obtained as
\begin{align}
\calH_\eff &= \calH_0 + \sum_{n > 0} \frac{[\calH_{-n},\calH_n]}{ n \omega} \nonumber \\
&=\sum_{\bm k \sigma} \tilde{\xi}_0 (\bm k) c_{\bm k \sigma} ^\dagger c_{\bm k \sigma} \nonumber\\
&+\sum_{\bm k \sigma \sigma'} (\alpha \bm \tilde{g}_0 (\bm k) \cdot \bm \sigma)_{\sigma \sigma'}
c_{\bm k \sigma} ^\dagger c_{\bm k \sigma'}+U \sum_i n_{i \uparrow} n_{i \downarrow}  \nonumber \\
&- \sum_{\bm k \sigma \sigma'} \mu_B \tilde{H}(\bm k)\sigma_z  c_{\bm k \sigma} ^\dagger c_{\bm k \sigma'},  \label{eff_model}
\end{align}
where
\begin{align}
\tilde{\xi}_0 ({\bm k})&=-2 t(J_0(A_x)\cos k_x+J_0(A_y)\cos k_y) \nonumber \\
&\qquad + 4t^\prime J_0 \left(\sqrt{A_x^2+A_y^2}\right)  \cos k_x \cos k_y - \mu, \\
\tilde{\bm g}_0 (\bm k)&= (-J_0(A_y)\sin k_y, J_0(A_x)\sin k_x, 0), \\
\tilde{H}(\bm k) &= - \frac{ 4 \alpha ^2 \calJ^2(A_x,A_y)}{\mu_B \omega}\cos k_x \cos k_y, \label{mag}\\
\calJ^2(A_x,A_y)&= \sum_{m = 0}  \frac{(-1)^{m}J_{2m+1} (A_x) J_{2m+1} (A_y)}{2m+1},
\end{align}
and $J_n(x)$ represents the $n$-th Bessel function.
We see two effects induced by the laser light.
The first one is so-called ``dynamical localization"\cite{Dunlap1986}.
With this effect, the hopping amplitude $t, t^\prime$ and the coupling constant $\alpha$ are renormalized by the 0-th Bessel function. This effect induces the deformation of Fermi surface, resulting in topological phase transitions as we mention below.    
The second one is ``laser-induced magnetic field". It causes the Zeeman splitting of which splitting-width varies in momentum space. It plays a crucial role in realizing TSC in this system.

\begin{figure}[tbp]
\includegraphics[width=9cm]{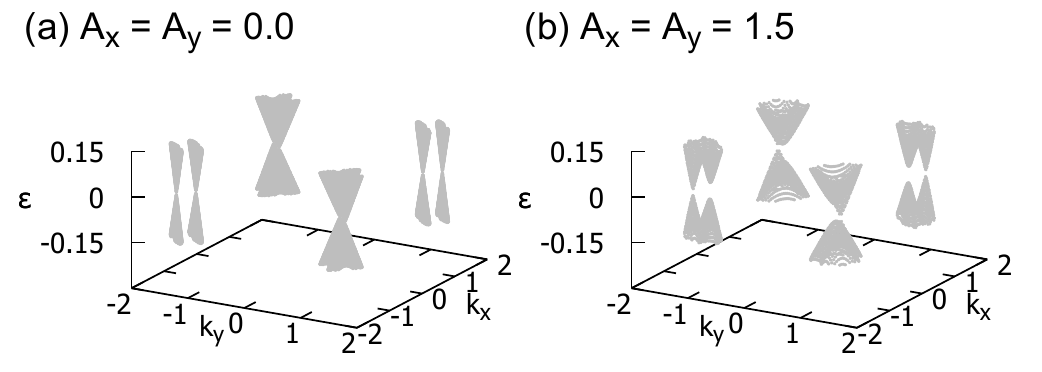}
\caption{Quasi-particle spectrum of the effective model (a) without laser-irradiation and (b) with laser-irradiation. The data are obtained by diagonalizing the BdG Hamiltonian (\ref{BdG}). We choose the parameters as $t=1.0$, $t^\prime=0.2$, $\alpha=0.3$, $\omega=0.4$, $\Delta_d = 0.4$, and $\Delta_p = 0.08$.}
\label{QP}
\end{figure}

Next we consider the Bogoliubov-de Gennes (BdG) Hamiltonian which describes superconducting states.
The order parameter of cuprate superconductors is known to be $d$-wave \cite{Yanase2003}.
However, in our system, the Rahba spin-orbit coupling breaks the inversion symmetry and thus $p$-wave pairing may be admixed with $d$-wave pairing \cite{Tada2009}.
Therefore we investigate the $D$+$p$-wave superconducting state by adopting a simple form $\Delta(\bm k)= i [\psi(\bm k)+ \bm d (\bm k) \cdot \bm \sigma]\sigma_y $
with $\psi(\bm k)= \Delta_d (\cos k_x - \cos k_y)$ and $\bm d (\bm k)= \Delta_p (\sin k_y, \sin k_x, 0)$. 
We assume that $|\Delta_d|$ is much larger than $|\Delta_p|$.
With the $D$+$p$ wave superconducting order parameter, we write down the BdG Hamiltonian as
$\calH_\mathrm{BdG} = 1/2 \sum_{\bm k} \Psi^\dagger_{\bm k} \calH(\bm k) \Psi_{\bm k}$,
where
\begin{align}
\calH(\bm k) &=
\begin{pmatrix}
\calH_N(\bm k) & \Delta (\bm k) \\
\Delta^\dagger(\bm k) & -\calH_N^T(- \bm k) 
\end{pmatrix}, \label{BdG}\\
\calH_N(\bm k)&= \tilde{\xi}_0 (\bm k) \sigma_0 + \alpha \bm \tilde{g}_0 (\bm k) \cdot \bm \sigma -\mu_B \tilde{H}(\bm k)\sigma_z, 
\end{align}
and $\Psi^\dagger_{\bm k} = (c^\dagger_{\bm k \uparrow}, c^\dagger_{\bm k \downarrow}, c_{-\bm k \uparrow}, c_{-\bm k \downarrow})$.
Without laser light ($A_x=A_y=0$), this model represents the original $D+p$ wave superconductor and thus it has point nodes (shown in Fig. \ref{QP} (a)).
With finite intensity of laser light ($A_x, A_y > 0$), the point nodes are gapped out  (shown in Fig. \ref{QP} (b)).
Later we show that the TSC is realized and the chiral Majorana edge modes appear in the laser-induced gap (shown in Fig. \ref{edge}).
This gap opening is caused by the laser-induced magnetic field (\ref{mag}). This term breaks the time-reversal symmetry and changes the symmetry class of the BdG Hamiltonian to class D. 

Before closing this section, we remark on differences from the case of usual magnetic field applied to cuprate superconductors. 
When the orbital depairing effect is neglected, this case is described by the model similar to ours, which has already been studied  by Yoshida and Yanase \cite{Yoshida2016, Daido2016}.
However, there are two important differences from their studies.
First, the laser-induced magnetic fields do not induce vortices in superconductors.
Usual magnetic fields induce vortices and easily suppress the superconducting states.
On the other hand, the synthetic magnetic field induced by the laser light leads to only the  Zeeman-type energy splitting (shown in Eq. (\ref{eff_model}) )
and oscillating gauge fields do not induce the vortices in superconductors.
This is an advantage in realizing the TSC. Second one is that the laser light induces hopping renormalization (dynamical localization). As we mention below,  the structure of superconducting gap is modified by the dynamical localization, and thus the topological states, which cannot be stabilized by usual magnetic fields, are realized.

\section{Topological properties}

\subsection{Chern number and Phase diagram}

As we mentioned above, our BdG Hamiltonian belongs to the symmetry class D in two dimensions. Therefore, the gapped state of this model is specified by the Chern number $C$\cite{Thouless1982}
, which is defined by
\begin{align}
C = \frac{1}{2 \pi i} \int d \bm k \epsilon^{i j} \sum_{n \mathrm{: filled}} \partial_{k_i} \braket{u_n(\bm k)| \partial_{k_j} u_n(\bm k)}.
\end{align}

We calculate the Chern number by two methods.
The first one is an analytical method, which is proposed by Daido and Yanase \cite{Daido2016}.
Their derivation can be straightforwardly applied to our model. 
The analytic form is obtained as 
\begin{align}
C = \sum_{(\pm , \bm k_0)} \frac{1}{2} \mathrm{sgn} \left [ 
\frac{(\hat{\bm z} \times \nabla_{\bm k} E_\pm ) \cdot \nabla_{\bm k} (\psi \pm \bm d \cdot \hat{\bm g})}
{\mu_B (\tilde{H} \hat{\bm z}) \cdot (\hat{\bm g} \times \bm d)/ \alpha} \right]_{\bm k = \bm k_0}, \label{formula}
\end{align}
where $\hat{\bm z}$ is a unit vector in the $z$-direction, $E_\pm = \tilde{\xi_0}(\bm k) \pm \alpha |\tilde{\bm g} (\bm k)|$ and $\hat{\bm g} = \tilde{\bm g} (\bm k) / | \tilde{\bm g} (\bm k)|$.
The summation is taken over all the gapped nodes $\bm k_0$ on Fermi surfaces of the $E_\pm$ bands. The energy spectrum is written as \cite{Daido2016}
\begin{align}
\mathcal{E}_{+} &= \pm \sqrt{E_+^2 + |  (\psi + \bm d \cdot \tilde{\bm g}) + i (\mu_B (\tilde{H} \hat{\bm z}) \cdot (\hat{\bm g} \times \bm d)/ \alpha |\tilde{\bm g}|) |^2} \label{band1}, \\
\mathcal{E}_{-} &= \pm \sqrt{E_-^2 + |  (\psi - \bm d \cdot \tilde{\bm g}) + i (\mu_B (\tilde{H} \hat{\bm z}) \cdot (\hat{\bm g} \times \bm d)/ \alpha |\tilde{\bm g}|) |^2} \label{band2}.
\end{align}
Therefore, the gap nodes appear in the absence of the layer light at $\bm k_0$ satisfying 
\begin{align}
E_\pm (\bm k_0) = \psi (\bm k_0) \pm \bm d (\bm k_0) \cdot \tilde{\bm g}(\bm k_0) =0. \label{condition}
\end{align}
The gapped nodes at $\bm k_0$ are intersections of a Fermi surface $E_\pm (\bm k)=0$ and 
zeros of order parameter $\psi (\bm k) \pm \bm d (\bm k) \cdot \tilde{\bm g (\bm k)}=0$.

\begin{figure}
\includegraphics[width=8.5cm]{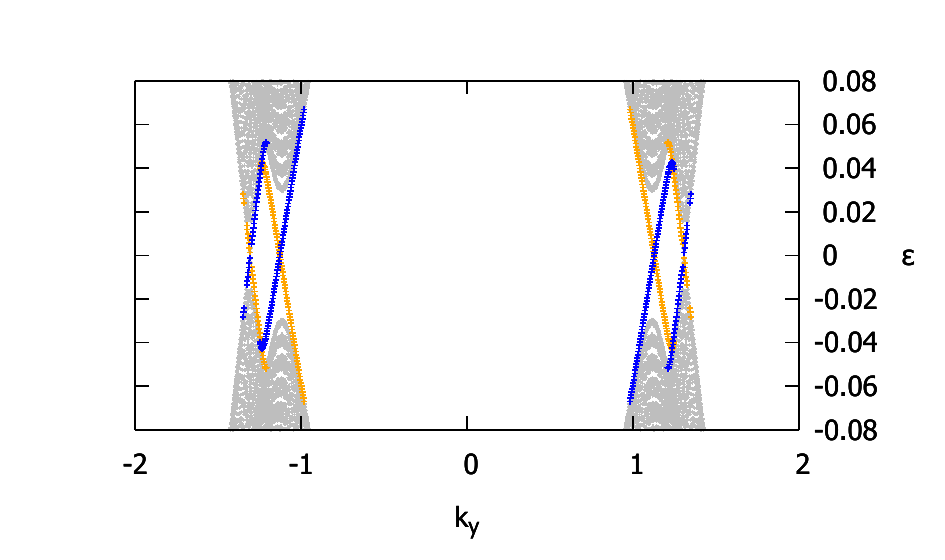}
\caption{Energy spectra in a ribbon-shaped system with open boundary conditions along the $x$-axis and periodic boundary conditions along the $y$-axis. The orange dots and blue dots show the Majorana mode localized at each side of edges, respectively. Four edge modes appear at each edge, corresponding to the Chern number $C=4$. We choose the parameters as $t=1.0$, $t^\prime=0.2$, $\alpha=0.3$, $\omega=0.4$, $\Delta_d = 0.4$, and $\Delta_p = 0.08$. The filling $n$ is $0.8$ ($n=1$ represents half-filling). }
\label{edge}
\end{figure}

\begin{figure}
\includegraphics[width=7cm]{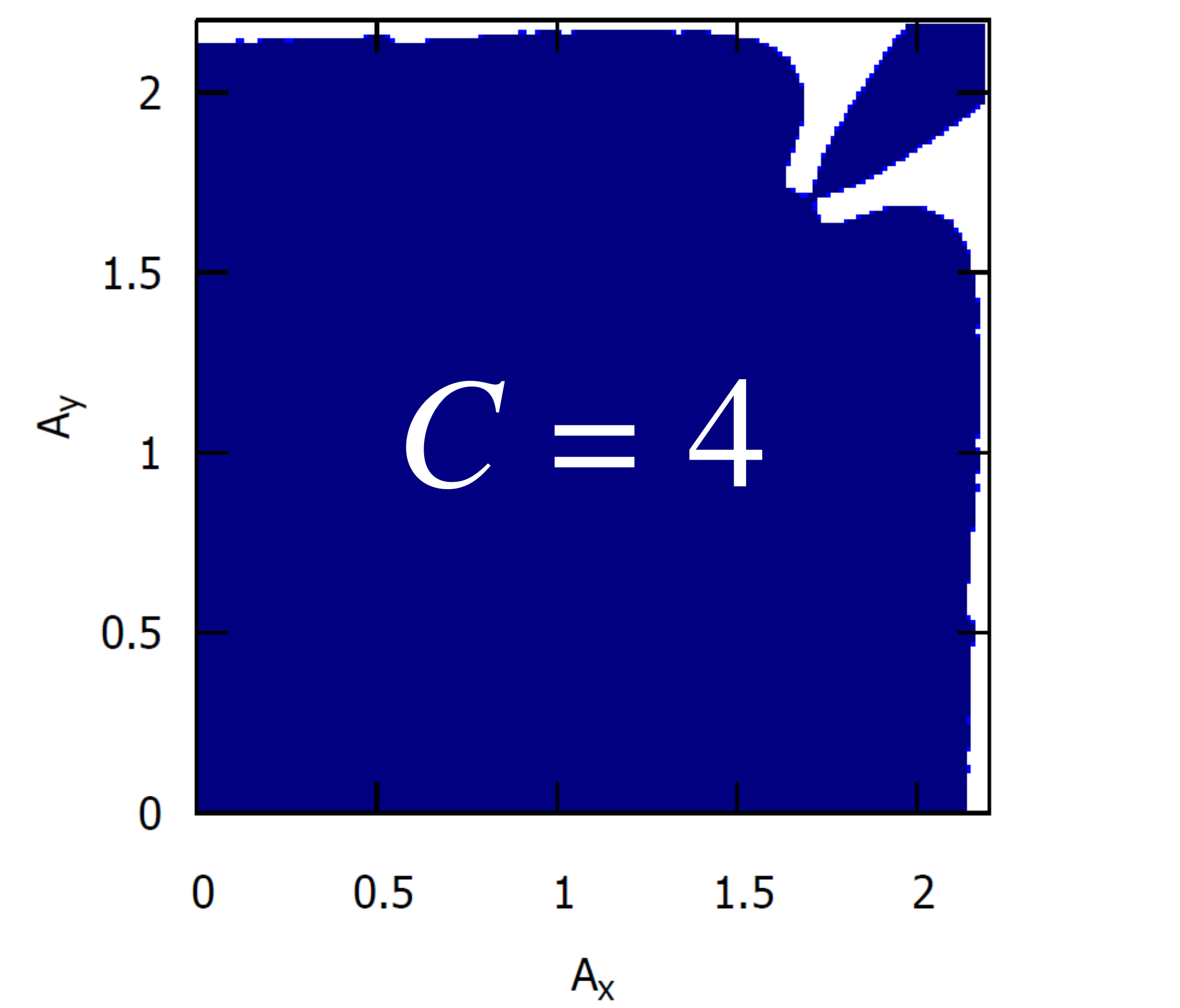}
\caption{Topological phase diagram of the laser-irradiated superconducting cuprate thin films. Color plot shows numerically calculated Chern numbers for each $(A_x, A_y)$ point. The white region shows a topologically trivial phase ($C=0$) and the blue region corresponds to a topologically nontrivial phases ($C=4$). We choose the parameters as $t=1.0$, $t^\prime=0.2$, $\alpha=0.3$, $\omega=36.0$, $\Delta_d = 0.05$, and $\Delta_p = 0.01$. }
\label{phase_diagram}
\end{figure}

Based on the analytic formula (\ref{formula}), 
we can evaluate the Chern number with counting the contribution from gapped nodes.
Each gapped node gives a contribution $+1/2$ or $-1/2$.
Sign of each contribution can be estimated as follows.
First we define the direction parallel to the Fermi surface of $E_\pm (\bm k)$ bands as 
$\hat{k}_{\pm} =  \hat{\bm z} \times \nabla_{\bm k} E_\pm / |\hat{\bm z} \times \nabla_{\bm k} E_\pm|$. 
Next we check the change of the sign of $\psi \pm \bm d \cdot \hat{\bm g} / (\mu_B (\tilde{H} \hat{\bm z}) \cdot (\hat{\bm g} \times \bm d)/ \alpha)$,
which is in the argument of the function of Eq. (\ref{formula}).
When it changes from negative to positive along $\hat{k}_{\pm}$ direction at gapped nodes, the contribution is $+1/2$, and vice versa.
Summing up all the contributions, we obtain the Chern number of the total bands.

This analytic formula is very useful for understanding the origin of the Chern number.
However, it is not convenient for systematic calculations in a broad range of parameters. 
Therefore, for systematic calculation in a broad range of parameters, we use another method to calculate the Chern number.
This is called Fukui-Hatsugai-Suzuki method \cite{Fukui2005},
which is an efficient numerical method to calculate the Chern number of the model defined on discretized momentum space. 
With this approach, we calculate the Chern number for each $(A_x, A_y)$ point and obtain the topological phase diagram which is shown in Fig. \ref{phase_diagram}.
In the phase diagram, the number of electrons is fixed by tuning the chemical potential $\mu$.
In some regions, the Chern number is finite, implying the TSCs. 
In the following subsections, Secs. IVB and IVC, we clarify the nature of the superconducting phases in the low intensity region ($A_x, A_y \lesssim 1.5$) and in the high intensity region ($A_x, A_y \gtrsim 1.5$), respectively.

\subsection{Weak intensity region}

In the weak intensity region, we find that a TSC specified by $C=4$ is realized in a broad range of parameters.
Even with infinitesimally weak intensity of laser light, 
the TSC is realized, and thus it is possible to experimentally realize TSCs with relatively weak laser light. 
This is one of the main results of this study. 
The energy spectrum of the ribbon-shaped system is shown in Fig.(\ref{edge}).
As expected from the bulk-edge correspondence, four chiral Majorana modes appear near the edge of the system. 
The number of chiral Majorana modes corresponds to the Chern number $C=4$.

\begin{figure}
\includegraphics[width=8cm]{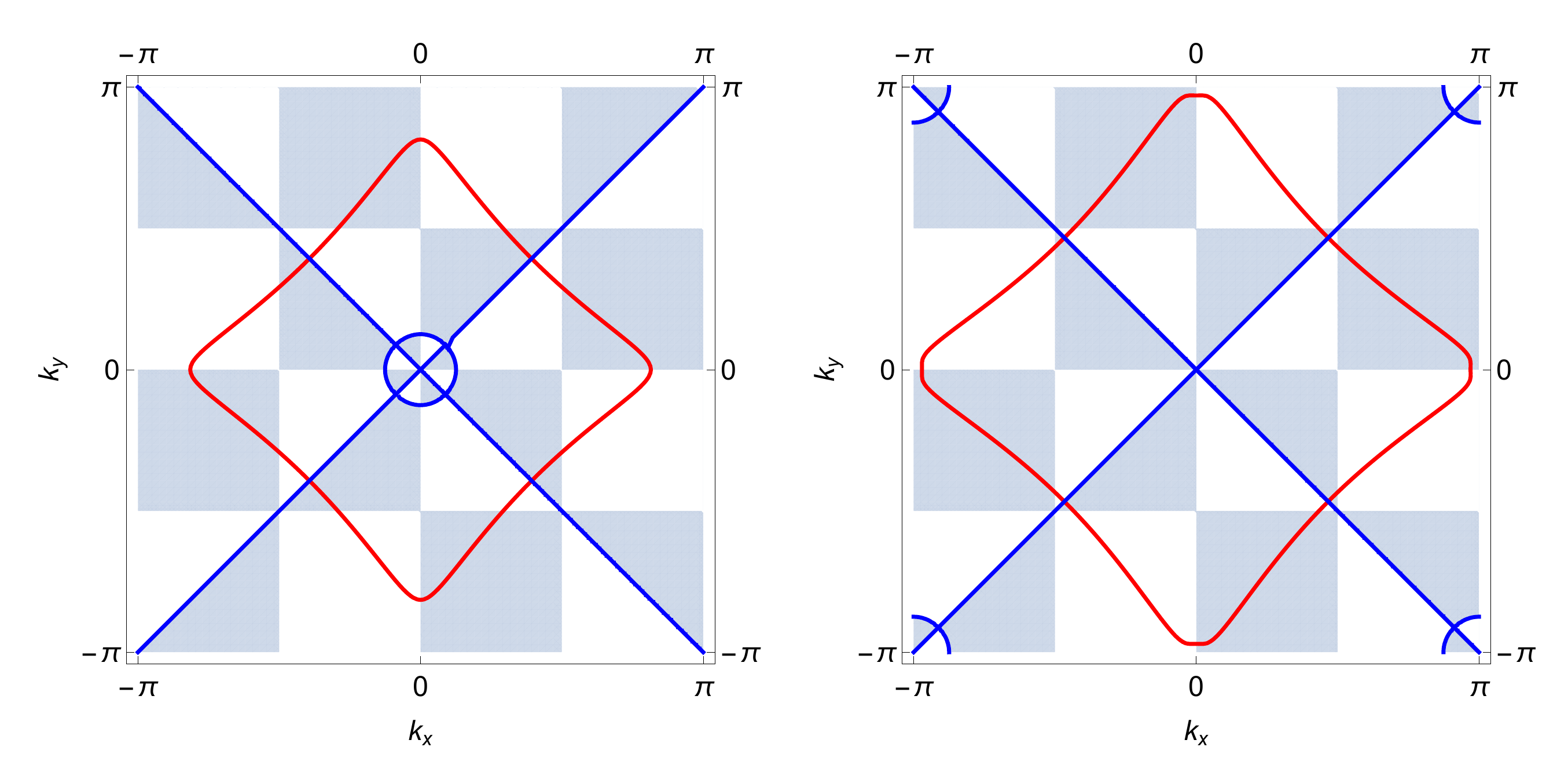}
\caption{Fermi surfaces (red lines) and zeros of superconducting gap (blue lines) with weak intensity of laser light $(A_x, A_y)=(0.1,0.1)$.
The $E_+(\bm k)$ band is shown in the left panel and the $E_-(\bm k)$ band is shown in the right panel.
The shaded region represents $(\psi \pm \bm d \cdot \hat{\bm g}) / (\mu_B (\tilde{H} \hat{\bm z}) \cdot (\hat{\bm g} \times \bm d)/ \alpha) > 0$. We use the same parameters as in the phase diagram, Fig.4 .}
\label{low}
\end{figure}

The reason why the $C=4$ phase is realized is understood with the analytic formula (\ref{formula}).
As we mentioned above, the Chern number is determined by the gapped nodes, which are defined as 
intersections of a Fermi surface and zeros of order parameters (they are shown in Fig. \ref{low}).
Dividing the contributions from two Fermi surfaces of the $E_+(\bm k)=0$ and $E_-(\bm k)=0$ bands, we write as $C=C_+ + C_-$.

We here evaluate $C_\pm$.  
In the case of circularly polarized laser ($A_x = A_y$), all of the four nodes on each Fermi surface are crystallographically equivalent since they are transformed by the four-fold rotation. 
Therefore, they give the same contributions \cite{Daido2016} and thus $C_\pm$ must be either 2 or -2.
Moreover, the superconducting gaps of each Fermi surface can be adiabatically deformed to each other without closing the gap, 
and thus the contributions to the Chern number is equivalent.
Therefore $C_+= C_-$ and we conclude that the Chern number is either 4 or -4. 
With the procedure mentioned in Sec. IVA, we find that the $C=4$ phase is realized. 
In the case of elliptically polarized laser($A_x \neq A_y$), the superconducting gap can be adiabatically deformed to that in the case of circularly polarized laser,
and thus the Chern number is not changed. Indeed, the $C=4$ phase is realized in a broad parameter range of the laser light as shown in Fig. \ref{phase_diagram}. 

At the end of this subsection, we discuss the effect of laser light on the superconducting order parameter.
Though the superconducting order parameter is assumed to be $D+p$ wave in this study, 
the nature of the superconducting order can be changed by laser light
through two effects, the paramagnetic effect and the deformation of the Fermi surface. 
However, the laser-induced magnetic field is very small and its effect on the superconducting order parameter is negligible. In the weak intensity region, deformation of Fermi surface is also small.
Therefore the assumption of the $D+p$-wave superconducting order is valid in this region.

\subsection{Strong intensity region}

\begin{figure}
\includegraphics[width=8cm]{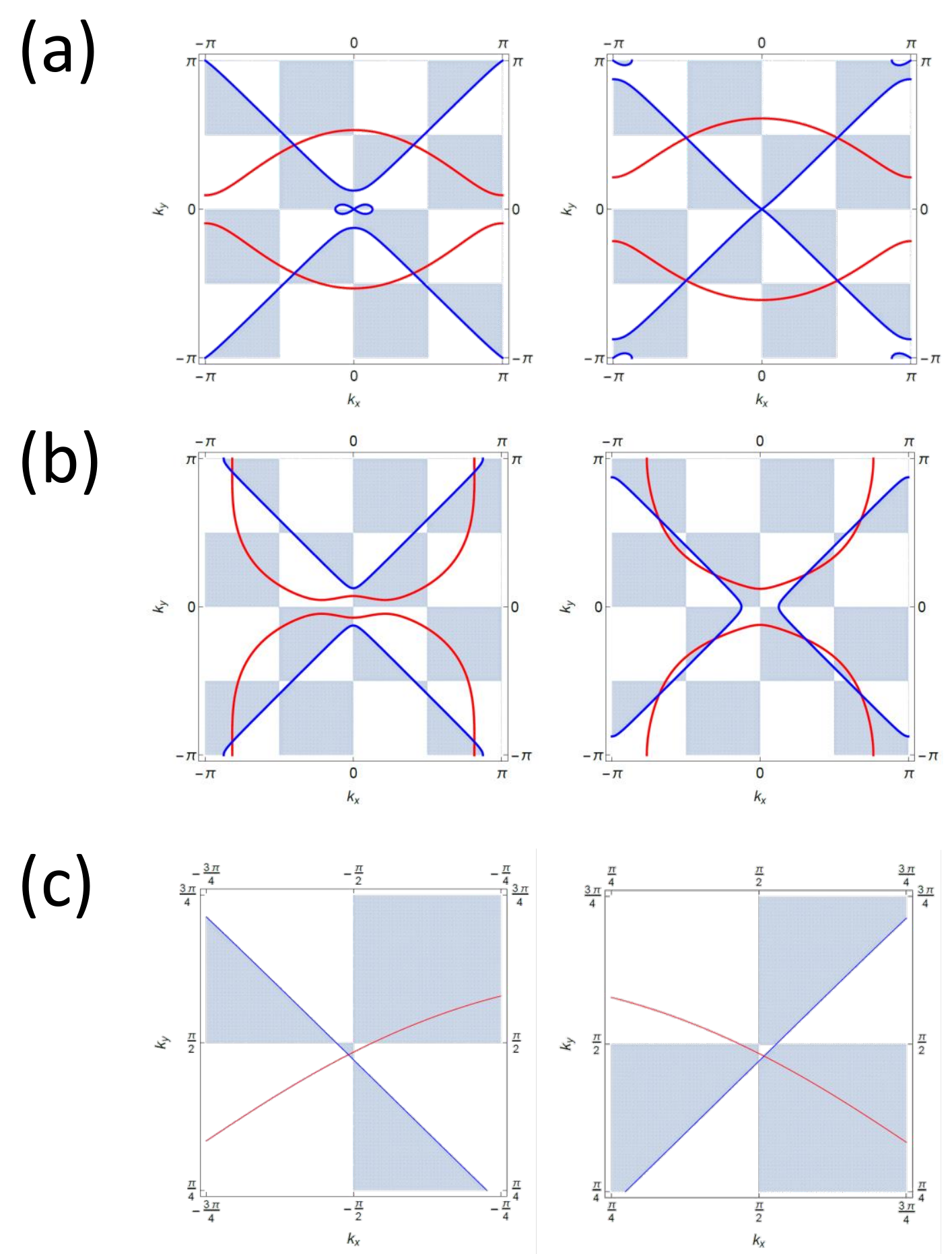}
\caption{(a, b) Fermi surfaces (red lines) and zeros of superconducting gap (blue lines) with strong intensity of laser light (a) $(A_x, A_y)=(2.1,1.8)$ and (b) $(A_x, A_y)=(2.8, 2.1)$.
The $E_+(\bm k)$ band is shown in the left panel and the $E_-(\bm k)$ band is shown in the right panel.
The shaded region represents $(\psi \pm \bm d \cdot \hat{\bm g}) / (\mu_B (\tilde{H} \hat{\bm z}) \cdot (\hat{\bm g} \times \bm d)/ \alpha) > 0$. We use the same parameters as in the phase diagram, Fig.4. (c) Enlarged figures of the right panel of Fig. (6.a). The detail of the structure near the gapped nodal points is shown.}\label{high}
\end{figure}

Even with strong intensity of laser light, the system shows the topologically non-trivial phase with $C = 4$.
Furthermore, under elliptic light  ($A_x \neq A_y$), different topological phases, such as $C= 0, 2$ and $-2$ are realized (The phases of $C=2$ and $-2$ appear out of the region of Fig. \ref{phase_diagram}). For example, the phase with $C=2$ appear in a finite region around the point $(A_x, A_y) = ( 2.8, 2.1)$, which is shown in Fig. \ref{high}(b). The appearance of $C= 0, 2$ and $-2$  reflects the fact that the rotation symmetry of the superconducting gap is reduced from four-fold to two-fold under the elliptic light.

The reason why these non-trivial Chern numbers are realized can also be understood with the analytic formula (\ref{formula}). 
Among them, we explain the phase of $C=0$ and $C=2$.
In Figs. \ref{high} (a) and \ref{high} (c), we show the case of $C=0$.
We can see that the symmetry of superconducting gap is reduced due to the elliptic laser light, but the system still has two-fold rotational symmetry.
By this symmetry, $C_\pm$ is restricted to $2, 0$ or $-2$. 
As we mentioned above, the sign of each contribution can be estimated from the sign of $\psi \pm \bm d \cdot \hat{\bm g} / (\mu_B (\tilde{H} \hat{\bm z}) \cdot (\hat{\bm g} \times \bm d)/ \alpha)$, which is shown in Fig. \ref{high} (a) by shading. 
The change of the sign is opposite between the $E_+ (\bm k)$ band and $E_- (\bm k)$ band.
Therefore $C_+$ and $C_-$ have opposite signs and thus $C=0$.
It is a topologically trivial state, which can be realized by strong laser irradiation.

Next we discuss the $C=2$ phase in Fig. \ref{high} (b).
The figure shows twelve nodes (four nodes in the left panel and eight nodes in the right panel).
Due to two-fold rotational symmetry, $C_+$ is limited to $2, 0$ or $-2$ and $C_-$ must be $4, 0$ or $-4$.
Estimating the contribution from each node, we obtain $C_+ = 2$ and $C_-=0$ and thus the Chern number $C$ is equal to $2$. Owing to numerical difficulties, the global phase diagram including the phases of $C= 2$ and $-2$ is not shown. However, we have confirmed that the phases of $C= 2$ and $-2$ appear in a finite region of the phase diagram as we mentioned above.

Finally, we remark on the effects of strong laser light on the superconducting order.
As we mentioned at the end of Sec. IVB, there are two laser-induced effects, paramagnetic effect and deformation of Fermi surfaces. 
Even in the strong intensity region, laser-induced magnetic field is still small for cuprates. 
However, the Fermi surfaces are drastically deformed and thus it is possible that the superconducting order is modified. 
However, the present system  is very likely to remain topologically-nontrivial as long as the nodal spin-singlet component is dominant\cite{Daido2016}. Therefore, we expect TSCs even when the superconducting gap is more or less deformed from the original $D+p$-wave one.

\section{Experimental conditions}
Our proposal has two advantages in experimentally realizing TSCs.
First one is that the topologically-non trivial states can be realized by infinitesimal intensity of laser light as we mentioned above. Indeed, any fine tuning of parameters is not required for TSCs. Second one is that the laser-induced magnetic field gives rise to only the Zeeman effect (paramagnetic effect) and thus does not induce vortices. 
Because of these advanteges, a laser-irradiated cuprate thin film is a promising candidate for TSCs.
In the following, we discuss the experimental conditions about materials, laser light and experimental methods. 

Concerning candidate materials, a cuprate superconductor has advantages
since its superconducting state is expected to be robust to the perturbations of laser light due to its high critical temperature.
With a slight modification (e.g. the form of $\xi (\bm k)$), our calculation can be applied to any $d$-wave superconductors with Rashba spin-orbit coupling.
Thus some of the heavy fermion superconductors, such as $\mathrm{CeCoIn_5}$ \cite{Shimozawa2014}, are also a candidate material for laser-induced topological superconductors.

In our model, the laser light is characterized by its frequency and intensity. 
Strictly speaking, the frequency must be sufficiently high and off-resonant 
since our calculation is based on the high-frequency expansion.
Within the scope of our calculation, frequency must be higher than the band width 8$t$. 
In solid state systems, there are so many unoccupied bands above Fermi energy, then we have to choose an appropriate frequency so as to make it off-resonant. 
On the other hand, even if the frequency is small, the signatures of the effective Hamiltonian may remain. 
For example, even by the low frequency laser light, a nodal point of Dirac cone is gapped out \cite{Wang2013}, which is the behavior analogous to the case of off-resonant driving.

Regarding the intensity of the laser light, 
the topological superconducting states ($C=4$) can be realized by infinitesimal intensity,
but weak intensity laser opens only a small gap, which is difficult to detect.
The largest energy gap is realized at $A_x=A_y \sim 1.84$, where the 1-st Bessel function takes a maximum value($\sim 0.58$).
The energy gap is estimated with eqs. (\ref{band1}) and (\ref{band2}) as,
\begin{align}
\left| \frac{\mu_B (\tilde{H}(\bm k) \hat{\bm z}) \cdot (\hat{\bm g}(\bm k) \times \bm d(\bm k))}{\alpha \tilde{\bm g}(\bm k)}\right|_{\bm k = \bm k_0} \sim 1.35 \frac{\alpha^2}{\omega} \frac{\Delta_p}{\tilde{\alpha}},
\end{align}
where the renormalized coupling constant is $\tilde{\alpha}=(J_0(A_x)^2+J_0(A_y)^2)^{1/2} \alpha$.
The amplitude of admixed $p$-wave component $\Delta_p$ is estimated as $\Delta_p \sim \Delta_d \tilde{\alpha} / E_F$ \cite{NCSC_book, Fujimoto2007} and thus the superconducting gap is evaluated as $0.1\mathrm{meV}$ when $\alpha=0.1\mathrm{eV}$, $\omega=10 \mathrm{eV}$ and $\Delta_d / E_F= 0.1$ are adopted. In this case, the laser intensity corresponds to the electric field amplitude $E \sim 1 \mathrm{GV/cm}$. 
In order to realize a large gap, we should prepare a system which has large $\alpha$ and 
apply laser with intermediate frequency as long as the high-frequency expansion is appropriate and the off-resonant condition is satisfied.

To experimentally probe the phenomena proposed here, we need to use a method suitable for observing transient phenomena
because we may use pulse laser in order to obtain strong intensity of laser light. Then, the TSC is realized only when laser light is applied. 
The most promising experimental tool is Tr-ARPES. The nodal structure of cuprate superconductors has been already observed in ARPES experiments\cite{Hashimoto2014}. 
Therefore, we believe that the gap-opening at the nodes will be detected with Tr-ARPES.  

\section{Summary and Outlook}
In this paper, we have suggested a possible way to realize TSCs with application of laser light to superconducting cuprate thin films \cite{Bollinger2011, Leng2011}. 
Using Floquet theory, we have analyzed the model of $d$-wave superconductors with Rashba spin-orbit coupling. We have derived an effective model and discussed its topological nature. The effective model includes the laser-induced magnetic fields, which make the system fully gapped and lead the system to the topologically non-trivial states characterized by Chern numbers. The laser-induced magnetic fields do not create vortices in the superconductors, and thus our proposal has an advantage in experimentally realizing TSCs in cuprates. 

We have chosen the laser frequency as being off-resonant in this study, but using different frequencies enables us to realize a different effective model. For example, it is proposed that a driving resonant to the interaction $U$ induces the correlated hopping and changes the superconducting nature\cite{Bukov2016}. Engineering the quantum states beyond off-resonant regime is an interesting direction for future work. 

\begin{acknowledgments}
This work is supported by JSPS KAKENHI (KAKENHI Grants 
No. JP16J05078, 
No. JP15H05884, 
No. JP15K05164, 
No. JP15H05745, 
No. JP16H00991, 
No. JP15H05855, 
and 
No. JP16K05501 
). KT thanks JSPS for the support from a Research Fellowship for Young Scientists.
\end{acknowledgments}

\appendix

\bibliographystyle{apsrev4-1}
\bibliography{ref.bib}

\end{document}